\documentclass[a4paper,twocolumn]{article}
\usepackage{lineno,hyperref}
\usepackage[T1]{fontenc}
\usepackage{graphicx,amsmath}
\usepackage{multirow}

\title{
	On the stability and existence of nitro-graphene, nitro-graphane, and nitro-graphene oxide
}

\author{Ruslan Yamaletdinov$^{*,1,2}$}
\date{
	$^1$ Boreskov Institute of Catalysis SB RAS,  Lavrentiev Ave. 5, Novosibirsk, Russia, 630090\\
	$^2$ Nikolaev Institute of Inorganic Chemistry	SB RAS, Lavrentiev Ave. 3, Novosibirsk, Russia, 630090\\
	$^*$ \textit{yamaletdinov@niic.nsc.ru}\\
	\today}

\begin{document}
	
	\twocolumn[
	\begin{@twocolumnfalse}
		\maketitle
		\begin{abstract}
			Here, the possibility of the existence of nitro-graphene derivatives with undisturbed graphene backbone is considered. Based on the first-principles calculation, it is shown that while NO$_2$ is unable to form a covalent bond with bare graphene, it becomes possible in some structures of graphane, fluorographene, and graphene oxide. This paper presents, among others, an analysis of the surrounding influence on the C-NO$_2$ bond strength. The author believes, that potential prospects of these materials and discussion about possible synthesizing routes may help further research on graphene-based materials.\\\\
		\end{abstract}
	\end{@twocolumnfalse}
	]

	\section*{Introduction}
	Functionalized graphene exhibits high potential for application in gas~\cite{Liu2017} and biosensors~\cite{Suvarnaphaet2017}; supercapacitors~\cite{Bakandritsos2019}, solar cells~\cite{Liu2015}, batteries~\cite{Lee2012}, and other energy materials; catalysis~\cite{Haag2014}, electromechanical devices~\cite{Panahi-Sarmad2019}, etc~\cite{Georgakilas2016}. This diversity is due to the broad range of properties tunable by functionalization type. It can be divided into covalent (chemical) and non-covalent (physical) functionalization, as interactions between functional additions and graphene~\cite{Jeon2011}. While there is an enormous number of composites and non-covalently bonded materials, the number of known covalently modified graphene materials with the original carbon backbone is limited~\cite{Chua2013}. Three most studied primary graphene derivatives with a covalent chemical bond are hydrogenated graphene (HG), fluorinated graphene (FG), and graphene oxide (GO). In turn, the nucleophilic substitution of functional groups in FG and GO can lead to a series of further derivatives~\cite{Whitener2015, Lai2011}.
	
	It was shown that various covalently bonded functional groups in GO, FG, and HG tend to form a unique pattern on graphene~\cite{Johns2013}. Various distributions lead to differences in physical properties, which introduces an additional degree of freedom in the adjustment of characteristics of such materials~\cite{Yamaletdinov2020a,Gargiulo2014}. Their functional groups patterns are mainly determined by energy parameters, such as the efficiency of the C-X interaction, violation of geometry, disruption of the graphene $\pi$-system, non-covalent interaction with the nearest neighbors, etc~\cite{Neek-Amal2015}. Based on this, the pattern of functionalized graphene with electron-withdrawing groups characterized by an extensive $\pi$ system should strongly differ from those known in FG, HG, GO, and therefore has a set of unique properties. NO$_2$ is one of the classic examples of such groups.
	
	There are many studies dedicated to the interaction of NO$_2$ molecule with graphene based material, in terms of application in gas-sensors (e.g.~\cite{Wang2016}). Such an application is possible due to a significant difference in the resistance of initial and  material bounded with the gas molecules. Numerous studies demonstrate the non-covalent interaction of the NO$_2$ with a plane of pristine or functionalized graphene with the binding energy 0-0.5 eV and distance 2-4 \r{A}~\cite{Maity2017}. Based on all these studies, it can be assumed that only edge atoms of graphene flakes, boundary atoms of sufficiently large vacancies can form a covalent C-NO$_2$ bond~\cite{You2017}. The presence of nitro groups is also reported after the modification of highly oxidized GO with a high concentration of -COOH groups and holes~\cite{Satheesh2014,Perez2020}. However, materials with such a damaged graphene backbone are poor conductors and subjected to rapid degeneration. In the same time, the rich chemistry of nitro compounds (especially a large number of possible reduced forms)\cite{Sassykova2019} encourage great potential in the RedOx applications of graphene with covalently bonded nitro groups.
	
	Recently, based on extensive density functional theory (DFT) dataset we conclude, that the single C-F binding energy is higher the more oppositely directed fluorine atoms in the first coordination sphere and the fewer fluorine atoms in the second coordination sphere~\cite{Yamaletdinov2020b} (see Fig.~\ref{fig:str_exmpl}). Inspired by this fact, here the possibility of covalent bonding of the -NO$_2$ group with graphene and its derivatives is considered.
	
	\section*{DFT calculations details}
\begin{figure}[!h]
		\begin{center}
			\includegraphics[width=0.35\columnwidth]{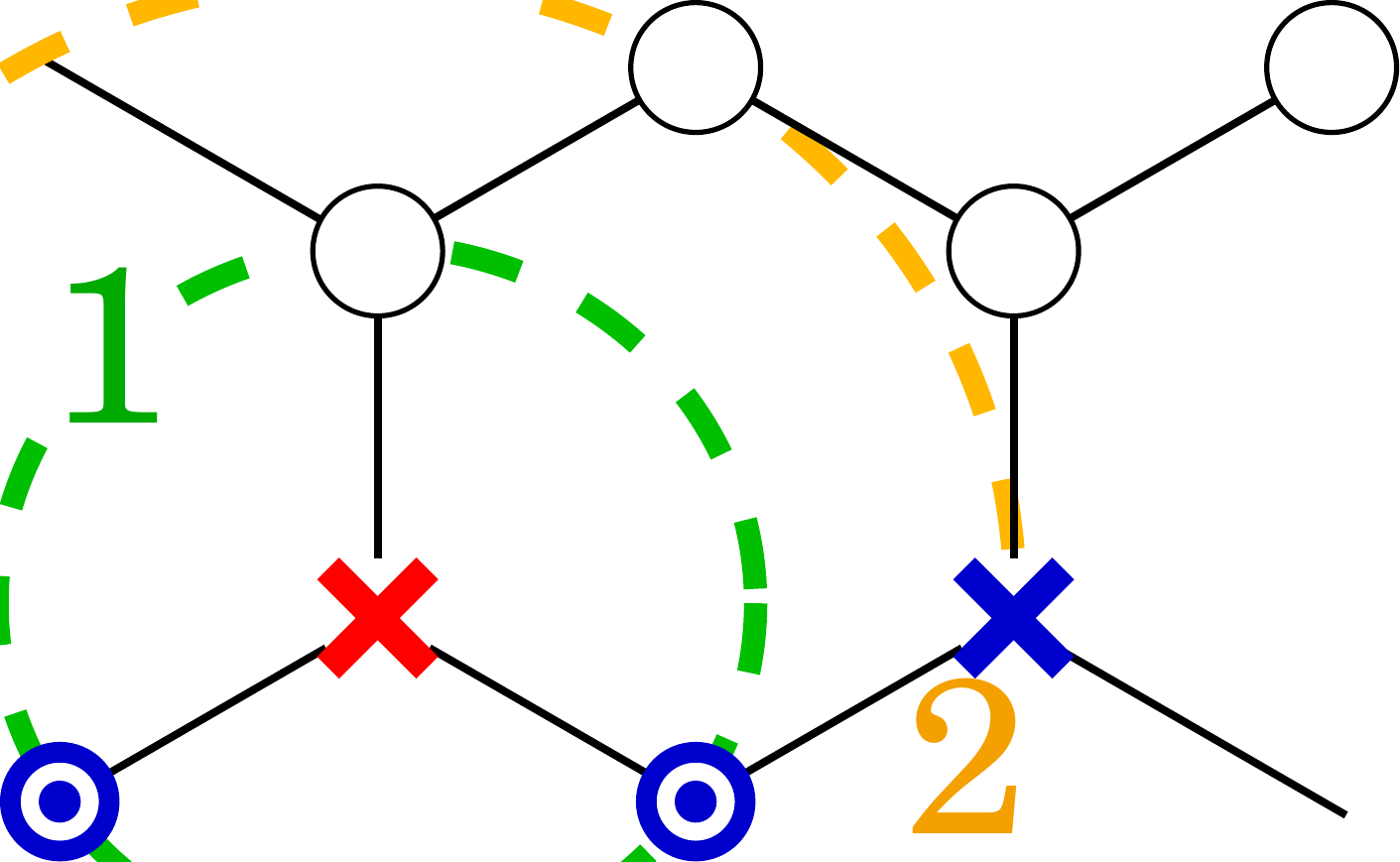}
		\end{center}
		\caption{An example of a unit cell. Green and orange dashed circles correspond to first and second coordination spheres. Blue colored atoms correspond to functional H- or F- or HO- groups. Red colored atom corresponds to -NO$_2$ group. Dot and cross notations correspond to up and down oriented functional group, respectively.}
		\label{fig:str_exmpl}
	\end{figure}

	DFT calculations were carried out using the QUANTUM-espresso toolkit~\cite{giannozzi2009quantum,giannozzi2017advanced} within Perdew, Burke, and Ernzerhof (PBE) exchange-correlation functional~\cite{Perdew1996}. PAW potentials with nonlinear core correction for each atom from \url{http://www.quantum-espresso.org} and corresponded recommended cutoff parameters were used. In this work were evaluated small unit cells of $2\times 2$ or $3\times 3$ primitive graphene unit cells (Fig.~\ref{fig:str_exmpl}, first column of table~\ref{tab:str}). 
	
	\section*{Results and discussions}
	An unsuccessful attempt to find any stable covalently bounded nitrographene C(NO$_2$)$_x$ was performed. Single NO$_2$ group, pairs, alternating chains C$_2$NO$_2$, fully nitrated CNO$_2$, and any distributed systems (such as C$_4$F in Ref.~\cite{Robinson2010}) do not converge in any stable covalently bonded structures in our calculations. It seems that large NO$_2$ groups mutually destabilize being even in the second coordination spheres of each other.

	\renewcommand{\arraystretch}{1.52}
	\begin{table}[!h]
		\caption{ C-NO$_2$ bond dissociation energy (in eV) and distances (in \r{A}) in various structures. C-X$_1$ and C-X$_2$ are the median C-X bond length in the first and second coordination sphere, respectively. The second line of GO is the length of the O-H bond of the first and second neighbors. }
		\begin{center}
			\begin{tabular}{|c|c|c|c|c|c|}
				\hline
				
				structure&&E$_\mathrm{CN}$&C-N&C-X$_1$&C-X$_2$\\\hline\hline
				\multirow{4}{*}{\includegraphics[width=0.2\columnwidth]{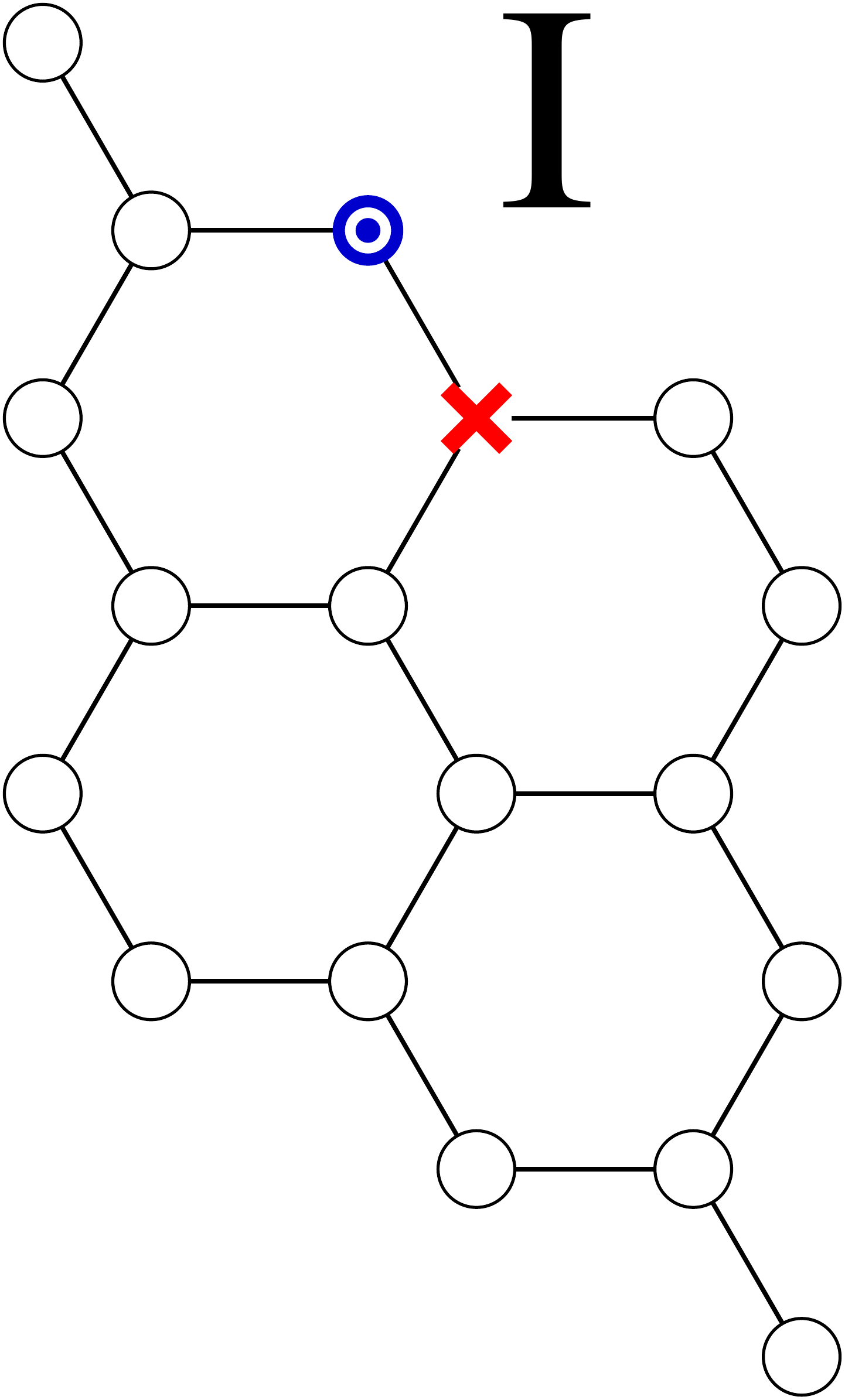}}&HG:&0.27&1.83&$1.13$&\\ \cline{2-6}
				&FG:&-0.25&1.74&$1.49$&\\ \cline{2-6}
				&\multirow{2}{*}{GO:}&-0.17&1.73&$1.48$&\\ \cline{3-6}
				&&&&$0.99$&\\ \hline \hline 
				
				\multirow{4}{*}{\includegraphics[width=0.2\columnwidth]{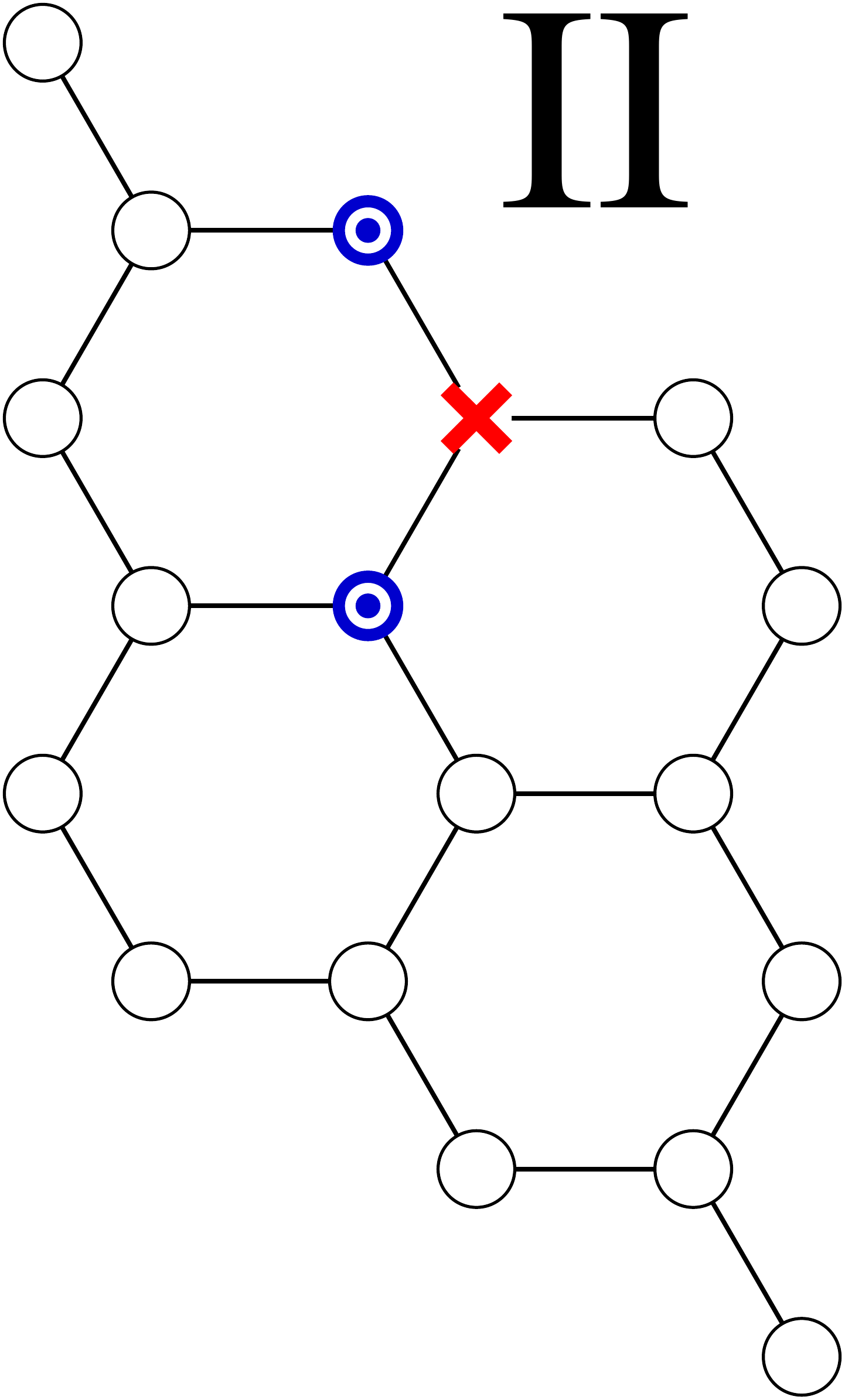}}&HG:&1.42&1.59&$1.12$&\\ \cline{2-6}
				&FG:&0.73&1.62&$1.46$&\\ \cline{2-6}
				&\multirow{2}{*}{GO:}&0.94&1.61&$1.48$&\\ \cline{3-6}
				&&&&$0.99$&\\ \hline \hline 
				
				\multirow{4}{*}{\includegraphics[width=0.2\columnwidth]{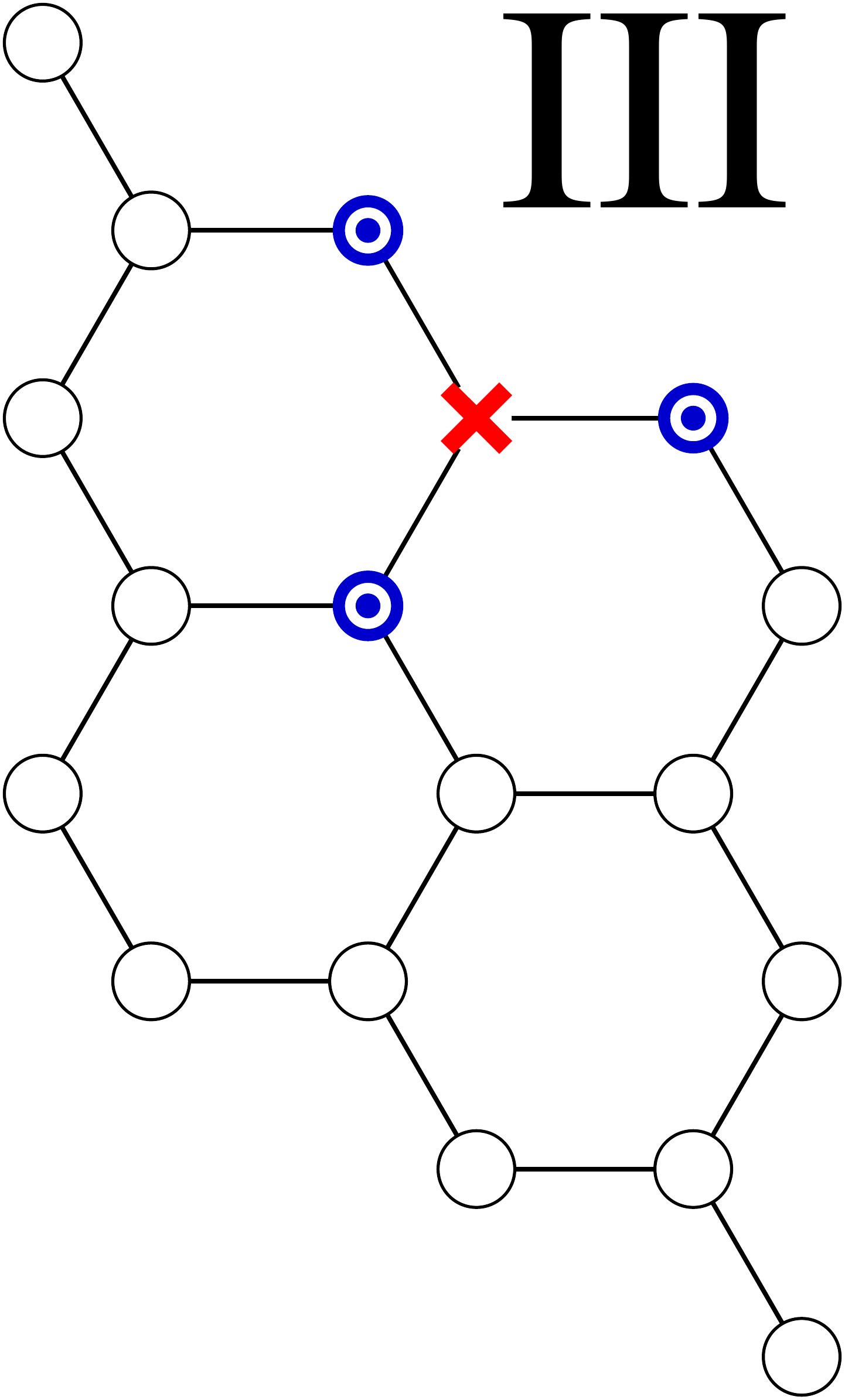}}&HG:&2.05&1.54&$1.12$&\\ \cline{2-6}
				&FG:&1.37&1.60&$1.45$&\\ \cline{2-6}
				&\multirow{2}{*}{GO:}&1.53&1.61&$1.46$&\\ \cline{3-6}
				&&&&$0.99$&\\ \hline \hline
				
				\multirow{4}{*}{\includegraphics[width=0.2\columnwidth]{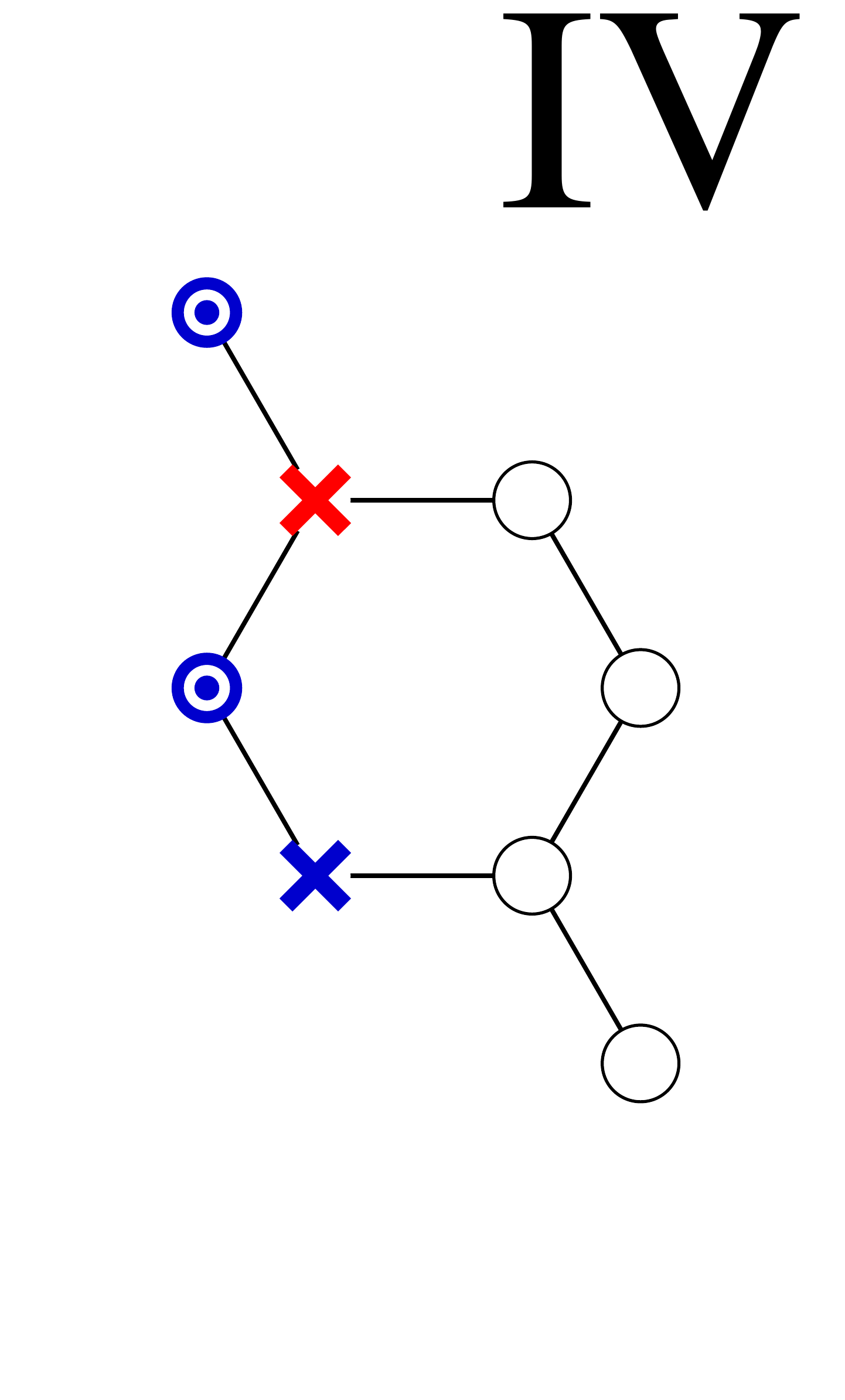}}&HG:&1.30&1.62&$1.11$&$1.10$\\ \cline{2-6}
				&FG:&0.88&1.68&$1.41$&$1.41$\\ \cline{2-6}
				&\multirow{2}{*}{GO:}&1.00&1.67&$1.45$&$1.43$\\ \cline{3-6}
				&&&&$1.00$&$0.98$\\ \hline \hline 
				
				\multirow{4}{*}{\includegraphics[width=0.2\columnwidth]{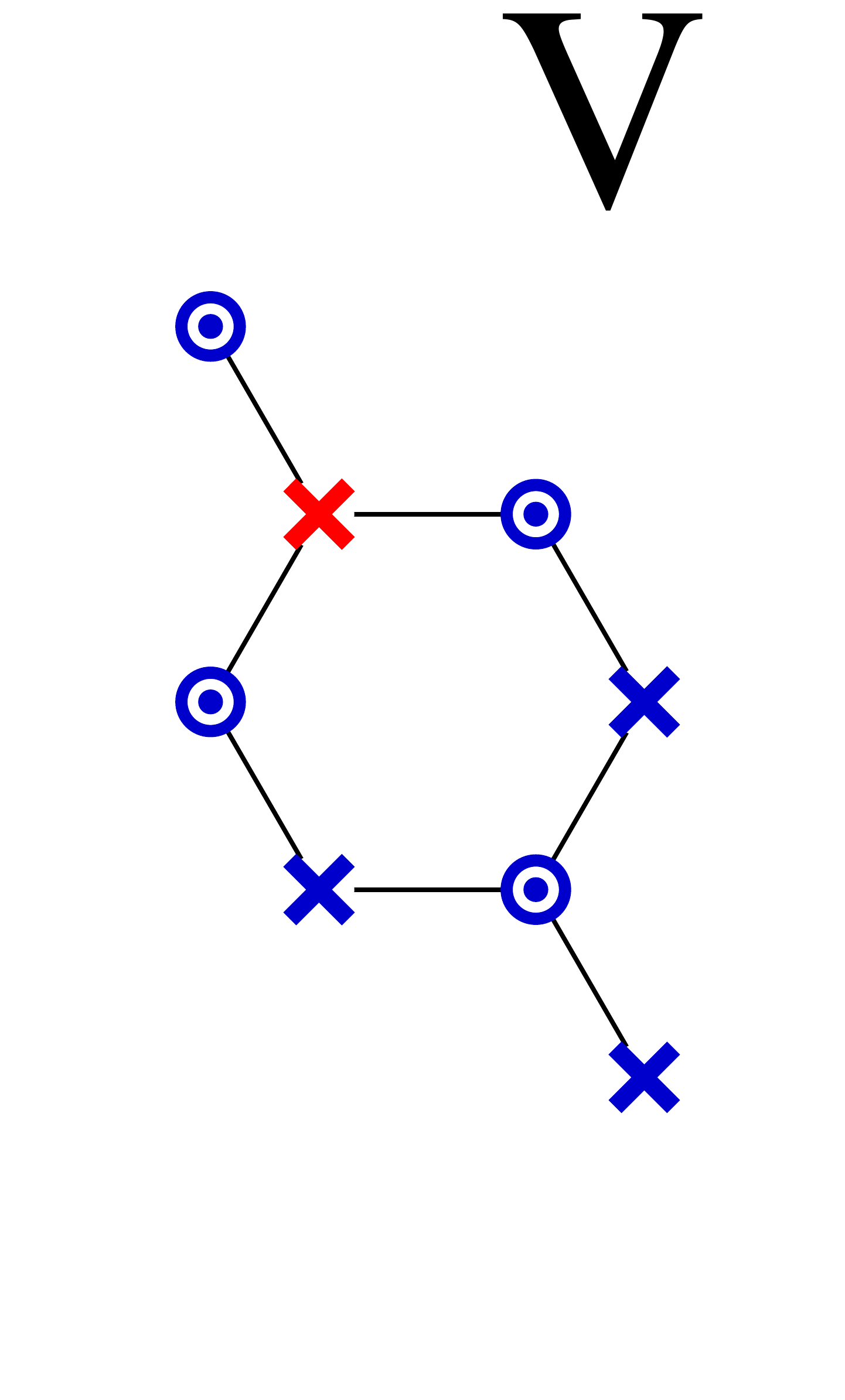}}&HG:&1.57&1.64&$1.11$&$1.09$\\ \cline{2-6}
				&FG:&-0.29&3.08&$1.41$&$1.38$\\ \cline{2-6}
				&\multirow{2}{*}{GO:}&-0.46&4.16&$1.45$&$1.40$\\ \cline{3-6}
				&&&&$1.00$&$0.99$\\ \hline  \hline 
				
				\multirow{4}{*}{\includegraphics[width=0.2\columnwidth]{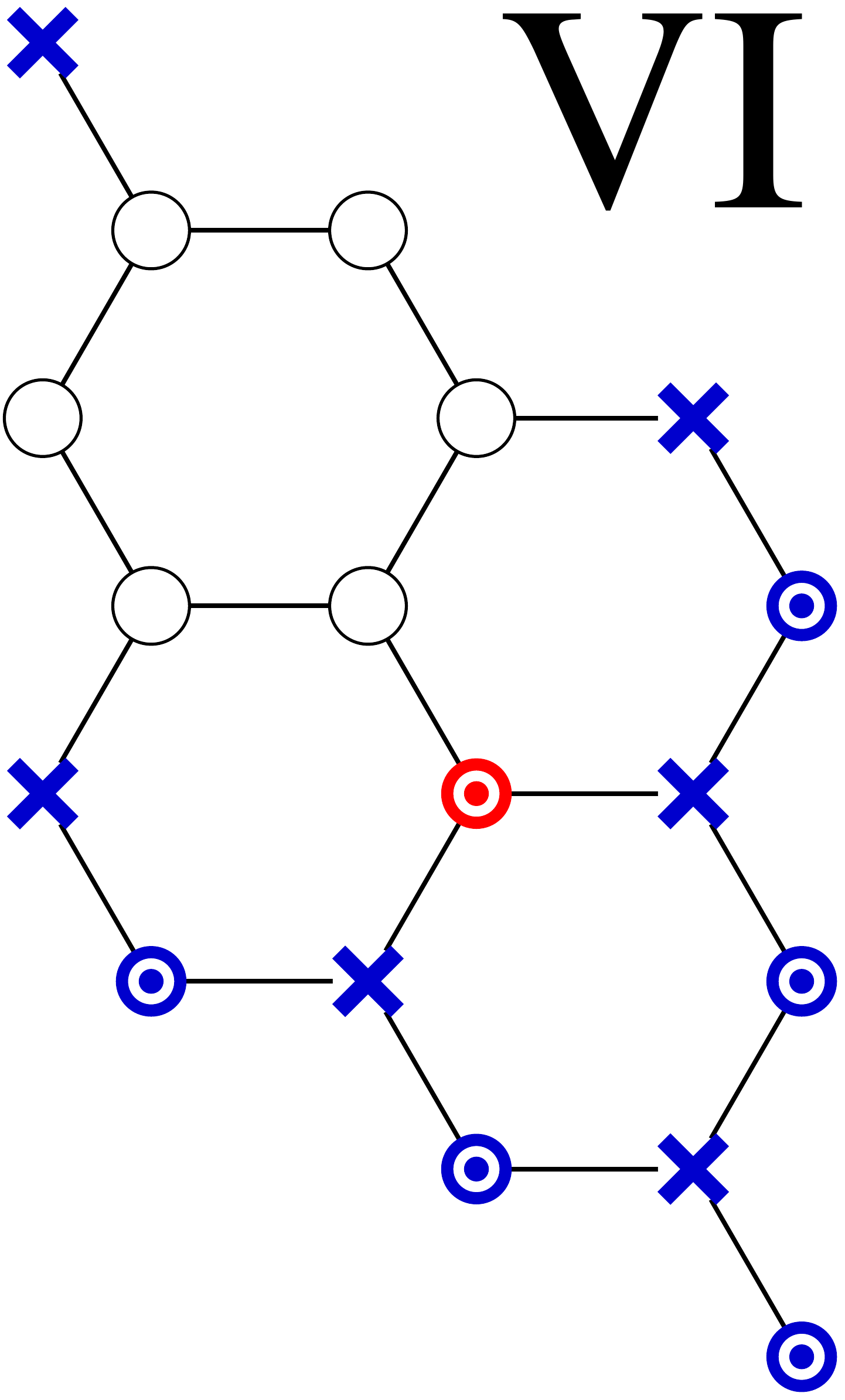}}&HG:&1.04&1.68&$1.11$&$1.10$\\ \cline{2-6}
				&FG:&0.06&2.31&$1.40$&$1.40$\\ \cline{2-6}
				&\multirow{2}{*}{GO:}&0.04&2.58&$1.44$&$1.43$\\ \cline{3-6}
				&&&&$1.00$&$1.00$\\ \hline

			\end{tabular}
		\end{center}
		\label{tab:str}
	\end{table}
	
	As is known, the presence of a neighbor in the first coordination sphere stabilizes H- in HG and F- in FG~\cite{Neek-Amal2015,Yamaletdinov2020b,Zhou2014}.  Based on this, simulation of structures with NO$_2$ group and one, two or three oppositely directed X (X = H, F, OH) in the first coordination sphere was carried out (see table~\ref{tab:str}, structures I-III). Obtained results is analyzed in terms of C-NO$_2$ bond dissociation energy and C-NO$_2$ and C-X$_{1,2}$ bonds length. C-NO$_2$ bond dissociation energy (E$_\mathrm{CN}$) is calculated as a difference between the energy of the XG structure with (E$_{\mathrm{XC-NO}_2}$) and without (E$_\mathrm{XC}$) the nitro group: E$_\mathrm{CN}= - [$E$_{\mathrm{XC-NO}_2} -$E$_\mathrm{XC}]$. It can be seen that the C-NO$_2$ and C-X binding increases as the number of neighbors in the first coordination sphere increases. Hydrogen has the strongest stabilizing effect, while hydroxyl and fluorine show almost the same influence, with a slight advantage for the hydroxyl group. Based on data obtained, we can conclude that structure I is the unlikely to be detected experimentally. It is difficult to conclude about an experimental observability of structures II and III, but the stability of XG in them without NO$_2$ is rather low~\cite{Johns2013}. 
	
	While neighbors in the first coordination sphere stabilize the structures, second neighbors obstruct it. Structures IV-VI were selected as one of the most stable for FG~\cite{Yamaletdinov2020b}. It is seen that the only for HG the covalent C-N bond can be formed in all described structures IV - VI. In the case of FG and GO, only the alternated chain (IV) is stable. Hydroxyl and fluorine again have the same effect on structure stability, but hydroxyl has the greatest destabilizing effect among neighbors in the second coordination sphere, which can be explained by the largest group size among X. At the same time, such a neighborhood decreases the length of the C-X bond in the second sphere, which can be interpreted as the C-X bonding enhancement. The more the destabilization of fluorine, the stronger the C-X bond. 
	
	\begin{figure}[!h]
		\includegraphics[width=\columnwidth]{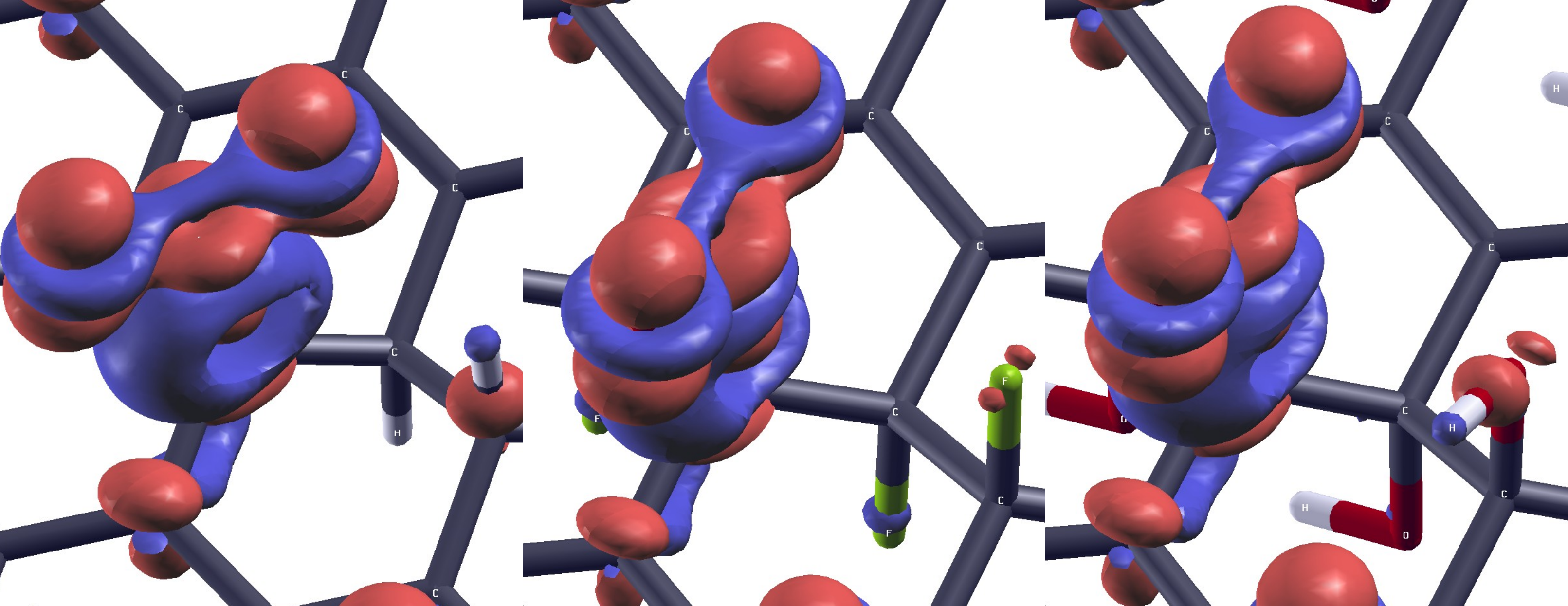}
		\caption{Electron density difference $\rho[\mathrm{C_nX_mNO_2}]-\rho[\mathrm{C_nX_m}]-\rho[\mathrm{NO_2}]$. From left to right NO$_2$ group on HG, FG, GO. Fragment from structure IV.}
		\label{fig:edens}
	\end{figure}
	
	The binding of NO$_2$ to the vacant carbon of structure IV leads to a noticeable redistribution of electron density (Fig.~\ref{fig:edens}). First of all, this binding leads to the polarization of NO$_2$ itself and to the destruction of the C = C $\pi$ bond. A strong negative inductive effect of NO$_2$ affects the first coordination sphere, which manifests itself in a slight decrease of electron density on X. The most interesting effect appears on the X of the second coordination sphere. It looks like electron density is shifted towards the $\sigma$(C-H) orbital for X = H, $p$(F) orbital for X = F, and $sp^2$(O)+$\sigma$(OH) for X = OH. Since NO$_2$ is the only available source of electrons, we can conclude that substituents in the second coordination sphere pull the electron density from NO$_2$ toward their bonding orbitals. The detailed mechanism of this effect requires further study.
	
\begin{figure}[!h]
		\includegraphics[width=\columnwidth]{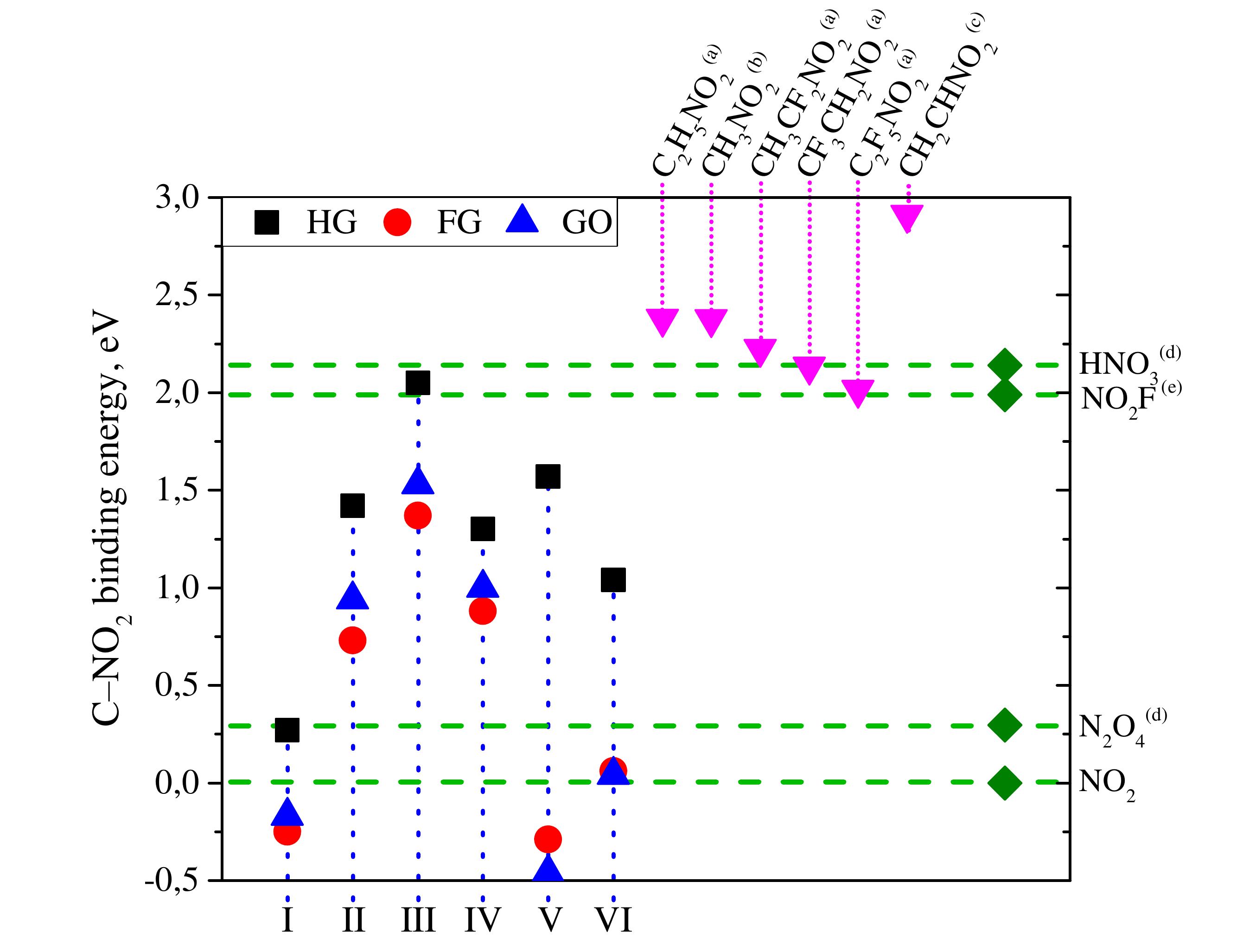}
		\tiny{\textit{a}-Ref.\cite{Khrapkovskii2004}; \textit{b}-Ref.\cite{Manaa1998}; \textit{c}-Ref.\cite{Shamov2009}; \textit{d}-Ref.\cite{NIST}; \textit{e}-Ref.\cite{Tschuikow-Roux1962}}
		\caption{C--NO$_2$ binding energies for various structures. Black, red and blue markers correspond to HG, FG, and GO structures from table~\ref{tab:str}. Purple triangles are there for comparing NO$_2$ bonding energy with some nitro-hydrocarbons and their derivatives. Green diamonds indicate the energy required to release one NO$_2$ molecule from the corresponding source.}
		\label{fig:energ}
	\end{figure}

	Since the first mentions of fluorographene and graphane, they are qualified as two-dimensional Teflon~\cite{Nair2010} and hydrocarbon~\cite{Sofo2007}, respectively. In this regard, the obtained C-NO$_2$ binding energy is compared with some nitroalkanes and fluoronitroalkanes (Fig.~\ref{fig:energ}). Figure~\ref{fig:energ} shows that the C-N binding energy of the structures considered in this work is mainly lower than in one-dimensional analogs. From this point of view, structure~III shows the greatest similarity with a difference of 0.33 eV (2.05 eV for structure~III vs 2.38 eV for nitromethane and nitroethane).
	
	Despite the relatively weak C-NO$_2$ bond, some of the considered structures are stable and, as expected, can be observed experimentally. However, no mention of this has been found. A possible explanation is the failure of the synthesis with conventional nitration methods and/or researchers may do not pay enough attention to the type of binding when studying the characteristics of gas sensors. For example, it has been reported that the UV-assisted sensing performance of HG is several times higher than that of GO, without any analysis of the binding type~\cite{Park2017}.
	
	So, under what conditions can nitro derivatives of graphene be synthesized, while a mixture of nitric and sulfuric acids is a commonly used method for graphene oxidizing~\cite{Jankovsky2017}? A comparison of the energy required for release one NO$_2$ molecule with C-NO$_2$ bond energy is carried out. The following precursors of NO$_2$ are considered (green dashed lines in Fig.~\ref{fig:energ}): HNO$_3$, NO$_2$F (for FG only), N$_2$O$_4$, NO$_2$. Gas-phase radical nitration can lead to nitro-graphane due to the great similarity of graphane with alkanes. However, the low stability of the C-NO$_2$ bond can prevent this even for HG. In the case of plasma synthesis of HG, FG and GO, a few number of unsaturated reactive centers remain on the structure~\cite{Yamaletdinov2020b}. These centers can form C-NO$_2$ bonds under a large excess of NO$_2$ or N$_2$O$_4$. However, this nitration method is unlikely to result in a structure with high NO$_2$ concentration. The most promising method is the binding of NO$_2$ during synthesis, when there are a large number of reactive centers. In the case of plasma modification of graphene, this can be achieved by adding NO$_2$ to the gas mixture. Liquid phase modification can be performed by autoclaving with a functionalizing agent (e.g. NO$_2$F for nitrofluorination) at room temperature with a N$_2$O$_4$ solvent.

	\section*{Conclusion}
	For now, it is difficult to predict the real-life stability of materials based on nitro-graphene derivatives.  Nevertheless, this study shows the possibility of their existence, as well as the first initial analysis of factors, influenced on nitro group stability, and proposes a possible synthesis approach. In any case, the ability to create graphene-based materials with such a RedOx variable group may have great potential in electrochemical applications.

	\section*{Acknowledgments}
	The reported study was funded by RFBR, project number 19-32-60012.
	
	\bibliographystyle{ieeetr}
	\bibliography{biblio}
\end{document}